\documentclass[aps,prd,nofootinbib,preprintnumbers,floatfix,twocolumn,superscriptaddress]{revtex4}

\usepackage{amsmath,amssymb,amsfonts,bm}
\usepackage{empheq}
\usepackage{physics}
\usepackage[svgnames,dvipsnames]{xcolor}
\usepackage{graphicx}
\definecolor{NewBlue}{rgb}{0.1, 0.1, 0.7}
\definecolor{NewRed}{rgb}{0.7, 0.1, 0.1}
\usepackage[colorlinks,
    linkcolor=Maroon,
    citecolor=NewBlue,
    urlcolor=NewRed]{hyperref}
\usepackage{cleveref}
\usepackage{lipsum}

\newcommand{\ie}{\textit{i.e.}}

\def\nn{\nonumber}

\def\l{\left}
\def\r{\right}
\def\d{{\rm d}}
\def\f{\frac}

\def\e{{\rm e}}
\def\O{\mathcal{O}}

\begin{document}

\title{Finite-time Unruh effect: Waiting for the transient effects to fade off}
\author{D. Jaffino Stargen}
\email{jaffinostargend@gmail.com}
\affiliation{Department of Physics, Global Academy of Technology,
Aditya Layout, RR Nagar, Bengaluru, Karnataka 560098, India}
\affiliation{Department of Mechanical Engineering, 
Massachusetts Institute of Technology,
Cambridge, MA 02139, USA}
\begin{abstract}
We investigate the transition probability rate of a Unruh-DeWitt (UD) detector interacting with
massless scalar field for a finite duration of proper time, $T$, of the detector. For a UD detector
moving at a uniform
acceleration, $a$, we explicitly show that the finite-time
transition probability rate can be written as a sum of purely thermal terms,
and non-thermal transient terms. While the thermal terms are independent of time, $T$, the non-thermal transient
terms depend on $(\Delta ET)$, $(aT)$, and $(\Delta E/a)$, where $\Delta E$ is the energy gap
of the detector. Particularly, the non-thermal terms
are oscillatory with respect to the variable $(\Delta ET)$, so that they
may be averaged out to be insignificant in the limit $\Delta ET \gg 1$, irrespective of the values of
$(aT)$ and $(\Delta E/a)$.
To quantify the contribution of non-thermal transient terms to the transition
probability rate of a uniformly accelerating detector, we introduce a parameter,
$\varepsilon_{\rm nt}$, called non-thermal parameter. Smaller the non-thermal parameter
is, smaller is the contribution due to non-thermal transient terms, when compared with the
purely thermal terms.
Demanding the contribution of non-thermal terms in the finite-time transition probability
rate to be negligibly small,
\ie, $\varepsilon_{\rm nt}=\delta\ll1$, we calculate the thermalization time -- the time required for the
detector to interact with the scalar field to arrive at the required non-thermality,
$\varepsilon_{\rm nt}=\delta$, so that the non-thermal terms to be negligibly small, and
the detector to be (almost) thermalized with the Unruh bath in its comoving frame.
Specifically, for small accelerations, $a\ll\Delta E$, we find the thermalization time, $\tau_{\rm th}$,
to be $\tau_{\rm th} \sim (\Delta E)^{-1} \times \e^{2\pi|\Delta E|/a}/\delta$;
and for large accelerations,
$a\gg\Delta E$, we find the thermalization time to be
$\tau_{\rm th} \sim (\Delta E)^{-1}/\delta$. We comment on the possibilities
of bringing down the exponentially large thermalization time at small accelerations, $a\ll\Delta E$.
\end{abstract}
\maketitle
\section{Introduction}
Unruh effect -- a robust prediction in quantum field theory in non-inertial frames -- refers to the
thermal character of Minkowski vacuum of a quantum field as "perceived" by a uniformly accelerating
observer \cite{Fulling1973,Davies1975,Unruh1976}. To be precise, the structure of Minkowski vacuum,
$\ket{0_{M}}$,
when restricted to a uniformly accelerating trajectory, takes the form of a thermal state, \ie,
$\ket{0_{M}} \bra{0_{M}} \propto \e^{-\beta {\hat H}}$, where ${\hat H}$ denotes the
Hamiltonian corresponding to the quantum field in the comoving frame of the accelerating observer,
and $\beta \equiv 2\pi/a$ (in natural units) signifies the inverse temperature of the Unruh bath,
with, $a\equiv\sqrt{-a^{\mu}a_{\mu}}$, denoting the proper acceleration of the observer
\cite{Takagi1986,Fulling1987,Wald1991,Crispino2008,Earman2011}.

An operational reformulation of Unruh effect involves a sufficiently localized two-level quantum
system, called Unruh-DeWitt (UD) detector, that is coupled linearly to the quantum field, and is
uniformly accelerating with a proper acceleration, $a$, with respect to inertial frames 
\cite{Unruh1976,DeWitt1979,Wald1984}. Due to the interaction between the detector and field,
the detector gets thermalized with the Unruh bath in its comoving frame. Particularly, it arrives
at a thermal state with inverse temperature, $\beta=2\pi/a$,
when it is allowed to interact with the field over a "sufficiently long period of time", irrespective of its
initial state \cite{Merkli2006,Garay2021}.

Though Unruh effect had been predicted roughly half a century back, an experimental realization testing
the prediction is still awaiting for fruition, primarily due to the experimental difficulties in achieving
extremely high accelerations, which is of the order of $a \gtrsim 10^{21}$ m/s$^2$
\cite{Takagi1986,Fulling1987,Wald1991,Crispino2008,Earman2011}. However,
various proposals were made that aim to enhance the effects due to Unruh bath
for accelerations
low enough, so that Unruh effect could be tested experimentally. Proposals employing optical cavities
\cite{Scully2003,Lochan2020,Stargen2022,Navdeep2022,Navdeep2023,Stargen2024,Majhi2024};
ultra-intense lasers \cite{ChenTajima1999,Habs2006,Habs2008};
Penning traps \cite{Rogers1988}; decay of accelerated protons \cite{Matsas2001}; radiation emission
in Bose-Einstein condensate \cite{Anglin2000,Reznik2008,Louko2020}; geometric phases
\cite{Mann2011,Ralph2022}; and
suitably selected Fock states \cite{Kalinski2005,Fuentes2010,Sudhir2022} are some of the notable
proposals made so far.

In standard theoretical treatments of Unruh effect
\cite{Takagi1986,Fulling1987,Wald1991,Crispino2008,Earman2011}, one assumes a uniformly
accelerating detector interacts with quantum field for an {\it infinitely} long period of time.
However, in
any realistic setup that aims to test the existence of Unruh effect,
a uniformly accelerating detector cannot interact with quantum field for arbitrarily
long periods of time, primarily due to the restrictions imposed by the feasibility of the experimental setup.
Therefore, from an experimental point of view, it is important to be aware of an estimate of the
time duration that a detector is required to be interacting with the quantum field for it to get
thermalized with the Unruh bath, and this is the primary purpose of this paper.

Finite-time UD detector -- a standard UD detector \cite{Unruh1976,DeWitt1979} that is allowed
to interact with quantum field
for a finite interval of time -- was investigated in various scenarios
\cite{Svaiter1992,Peres1993,Padmanabhan1996,Satz2008,Louko2012,Louko2014,Louko2016,Jenkinson2025,Moustos2019}.
Particularly, the transition probability of a finite-time UD detector in (3+1)-dimensions
was shown to be logarithmically divergent, if the interaction between the detector and quantum field
is instantaneously switched on and off \cite{Svaiter1992,Peres1993};
but the finite-time transition probability rate in (3+1)-dimensions was shown
to be finite, even if the detector is switched on and off instantaneously
\cite{Satz2008,Louko2012}. Since transition probability rate
is a meaningful physical observable for the case of
detector interacting with the field for {\it infinitely} long duration of time
\cite{Takagi1986,Fulling1987,Wald1991,Crispino2008,Earman2011}, it is convenient to consider
the same as the observable of interest for the case of finite-time UD detector too.
This allows one to estimate the time
duration that the finite-time detector requires to get it thermalized with the Unruh bath, particularly,
by comparing the finite-time response rate with the (purely thermal) infinite-time response rate.

This paper is organized as follows: In section-\ref{Sec.-II} we describe the setup by introducing
the relevant variables of interest, and calculate the general expression for finite-time
response rate of a detector interacting with massless scalar field.
In section-\ref{Sec.-III} and section-\ref{Sec.-IV}, we
calculate the finite-time response rate of an inertial, and a uniformly accelerating
detector, respectively. In section-\ref{Sec.-V}, we estimate the
time duration that a uniformly accelerating detector requires, so that the non-thermal
transient terms in the finite-time response rate becomes insignificantly small. Finally, we conclude
our results in section-\ref{Sec.-VI}.
\section{Finite-time response rate of a UD detector}\label{Sec.-II}
Consider a UD detector -- a sufficiently localized two-level quantum system,
with ground and excited states denoted as $\ket{g}$ and $\ket{e}$, respectively; and is
described by the Hamiltonian, ${\hat H}_{\rm UD}=\Delta E {\hat \sigma}_{z}/2$, with
${\hat \sigma}_{z}=(\ket{e}\bra{e}-\ket{g}\bra{g})$,
and $\Delta E$ denoting the energy gap of the detector -- interacting with the massless scalar field,
${\hat \phi}(x)$, and is allowed to move in a given spacetime trajectory, say, ${\tilde x}(\tau)$,
with $\tau$ denoting the proper time of the detector in its comoving frame.
The interaction between the scalar field, ${\hat \phi}(x)$, and the detector is described by the
interaction Lagrangian,
${\cal L}_{\rm int}=\lambda {\hat m}(\tau) {\hat \phi}[{\tilde x}(\tau)]$, where $\lambda$
is a small coupling constant, and ${\hat m}(\tau)$ denotes the monopole moment of the detector
\cite{Takagi1986,Fulling1987,Wald1991,Crispino2008,Earman2011}.

If the interaction between the detector and the quantum field is switched on and switched off
in a particular manner, which is captured by introducing a switching function, $\chi(\tau)$,
then the transition probability of the detector can be found to be
\cite{Takagi1986,Fulling1987,Wald1991,Crispino2008,Earman2011,Svaiter1992,Peres1993,
Padmanabhan1996,Satz2008,Louko2012,Louko2014,Louko2016,Jenkinson2025,Moustos2019}
\begin{align}
 {\cal P}(\Delta E) = \lambda^2 |\bra{e} {\hat m}(0) \ket{g}|^2 \times {\cal F}(\Delta E),
\end{align}
where
\begin{align}\label{eqn:Res-Gen}
 {\cal F}(\Delta E) &\equiv \int_{-\infty}^{\infty} d\tau \chi(\tau) \int_{-\infty}^{\infty} d\tau'
 \chi(\tau') \nn \\
 &\times e^{-i\Delta E(\tau-\tau')} {\cal W}(\tau,\tau'),
\end{align}
is known as the response of the detector, with ${\cal W}(\tau,\tau')
\equiv \langle0| {\hat \phi}[{\tilde x}(\tau)] {\hat \phi}[{\tilde x}(\tau')]|0\rangle$
denoting the Wightman function on a given trajectory, ${\tilde x}(\tau)$, of the detector.
The expression for Wightman function in cylindrical polar coordinates, $(\rho,\theta,z)$, can be written as
\begin{align}\label{eqn:Wightman}
 {\cal W}(x,x') &= \f{1}{(2\pi)^2} \sum_{m=-\infty}^{\infty}
 \int_0^\infty \d q~ q J_{m}(q\rho) J_{m}(q\rho') \nn \\
 &\times \int_{-\infty}^{\infty} \f{\d k_z}{2\omega_k}
 \e^{-i\omega_k (t-t')} \e^{im(\theta-\theta')} \e^{ik_z(z-z')},
\end{align}
where $J_{m}(z)$ denotes the Bessel function of the first kind \cite{Abramowitz1972,Gradshteyn2007}.

If the time duration, $T$, in which the detector is interacting with the quantum field
is finitely large, and the switching function, $\chi(\tau)$, satisfies $|\chi(\tau)| \approx 1$
within the time duration, $T$,
then the expression for detector response, ${\cal F}$, in Eq.(\ref{eqn:Res-Gen}) may be
approximated to
\begin{align}\label{eqn:Res-Gen_1}
 {\cal F}(\Delta E) \approx {\cal F}_{T}(\Delta E)
 &\equiv \int_{-T/2}^{T/2} d\tau \int_{-T/2}^{T/2} d\tau' \nn \\
 &\times e^{-i\Delta E(\tau-\tau')} {\cal W}(\tau,\tau').
\end{align}

If the detector moves along the integral curve of a Killing
vector field corresponding to the background Minkowski spacetime, then
the expression for response of the detector in Eq.(\ref{eqn:Res-Gen_1}) gets simplified to
\cite{Svaiter1992}
\begin{align}\label{eqn:response}
 {\cal F}_{T}(\Delta E) &= T\int_{-T}^{T} du e^{-i\Delta Eu} {\cal W}(u,0) \nn \\
 &- \int_{-T}^{T} du~{\rm sgn}(u)~u~e^{-i\Delta Eu} {\cal W}(u,0).
\end{align}
It is well known that the response rate of an infinite-time detector, which we denote as
${\dot {\cal F}}_{\infty}(\Delta E)$, is a meaningful observable for the detector
interacting with the field for an infinitely long period of time, \ie, $T\to\infty$
\cite{Takagi1986,Fulling1987,Wald1991,Crispino2008,Earman2011}.
Since we aim to estimate the time duration that a detector is required to interact
with the field, so that the detector reaches appreciably close to thermality, it is convenient to consider
the finite-time response rate, ${\dot {\cal F}}_{T}(\Delta E)$, as our observable of interest,
and we define the same as
\begin{align}
 {\dot {\cal F}}_{T}(\Delta E) \equiv \f{d{\cal F}_{T}(\Delta E)}{dT}.
\end{align}

Making use of the expression for finite-time detector response, ${\cal F}_{T}(\Delta E)$,
in Eq.(\ref{eqn:response}), we calculate the expression for finite-time response rate as
\cite{SuppMat}
\begin{align}\label{eqn:ResRate}
 {\dot {\cal F}}_{T}(\Delta E) = \int_{-T}^{T} du~ e^{-i\Delta Eu} {\cal W}(u,0).
\end{align}
Employing the above expression, we calculate the finite-time response rate for inertial and
uniformly accelerating detectors in the forthcoming sections.
\section{Finite-time response rate of an inertial detector}\label{Sec.-III}
Before considering the finite-time response rate of a uniformly accelerating detector,
it is instructive to consider the simplest case where the detector is moving along an inertial
trajectory with velocity, $v$, \ie,
${\tilde x}(\tau)=(t,\rho,\theta,z)=(\gamma\tau,\rho_{0},\theta_{0},\gamma v\tau)$,
where $\gamma \equiv 1/\sqrt{1-v^2}$,
$\rho_{0}$ and $\theta_{0}$ are constants that denote the radial and angular positions
of the detector in cylindrical polar coordinates. Making use of the information about the trajectory
of the detector in the expression for Wightman function, ${\cal W}(x,x')$, in Eq.(\ref{eqn:Wightman}),
one obtains \cite{Takagi1986,Fulling1987,Wald1991,Crispino2008,Earman2011}
\begin{align}
 {\cal W}(u,0) = -\f{1}{(2\pi)^2} \f{1}{(u-i\varepsilon)^2},
\end{align}
where $\varepsilon>0$. Substituting this in the expression for finite-time response rate,
${\dot {\cal F}_{T}}(\Delta E)$, in Eq.(\ref{eqn:ResRate}), one obtains
the expression for finite-time inertial response rate, ${\dot {\cal F}_{T}}^{(i)}(\Delta E)$,
as \cite{Svaiter1992,Peres1993,Padmanabhan1996,Satz2008}
\begin{align}\label{eqn:ResRateInertial}
 & {\dot {\cal F}_{T}}^{(i)}(\Delta E) = \f{\Delta E}{2\pi}
 \biggl\{-\Theta(-\Delta E) \nn \\
 &+ \f{\cos(\Delta ET)}{\pi \Delta ET}
 + \f{1}{\pi} {\rm Si}(\Delta ET)-\f{1}{2}{\rm sgn}(\Delta E)\biggr\}.
\end{align}
Note that the terms in the second line of the above expression for finite-time inertial
response rate are due to the
transient effects that originate from the {\it finite-time} interaction of the detector with
the (simple harmonic) field
modes. Interestingly, these transient effects vanish when the detector is allowed
to interact with the field for a sufficiently long period of (proper) time in its comoving frame.

If the detector is allowed to interact with the quantum field for a time
duration, $T$, that is
much larger than the time scale associated with the energy gap of the detector, \ie,
$(\Delta E)^{-1}$, then the expression for finite-time inertial response rate in
Eq.(\ref{eqn:ResRateInertial}) reduces to \cite{Svaiter1992,Peres1993,Padmanabhan1996,Satz2008}
\begin{align}
 {\dot {\cal F}^{(i)}_{T}}(\Delta E) &= \f{\Delta E}{2\pi} \biggl\{-\Theta(-\Delta E) \nn \\
 &- \f{\sin (\Delta ET)}{\pi (\Delta ET)^2} + \O\l(\f{1}{(\Delta ET)^3}\r)\biggr\}.
\end{align}
Therefore, in the limit $\Delta ET\to\infty$, the finite-time inertial response rate,
${\dot {\cal F}^{(i)}_{T}}(\Delta E)$, becomes
\cite{Takagi1986,Crispino2008,Earman2011,Svaiter1992,Peres1993,Padmanabhan1996,Satz2008}
\begin{align}
 \lim_{\Delta ET \to \infty} {\dot {\cal F}^{(i)}_{T}}(\Delta E)
 = -\f{\Delta E}{2\pi} \Theta(-\Delta E),
\end{align}
which showcases the fact that the transient effects in finite-time inertial detector
become insignificant when the
detector is allowed to interact with the quantum field for time durations, $T$, much larger
than the energy gap,
$\Delta E$, of the detector, \ie, $\Delta ET\gg1$. In other words, for the finite-time inertial
detector, the energy gap of the detector, $\Delta E$, sets the time scale after which the
transient effects start becoming insignificant.
\section{Finite-time response rate of a uniformly accelerating UD detector}\label{Sec.-IV}
Let us turn to the case of a UD detector that is on a uniformly
accelerating trajectory,
\ie, ${\tilde x}(\tau)=(t,\rho,\theta,z)=(a^{-1}\sinh a\tau,\rho_{0},\theta_{0},a^{-1}\cosh a\tau)$,
where $a\equiv(-a_{\mu}a^{\mu})^{1/2}$ is the proper acceleration of the detector, $\rho_{0}$
and $\theta_{0}$ are constants that denote the radial and angular positions of the detector in
cylindrical polar coordinates. Making use of the information about the trajectory of the
detector in the expression for Wightman function in Eq.(\ref{eqn:Wightman}),
and substituting it in the expression for finite-time response rate of the detector,
${\dot {\cal F}}_{T}(\Delta E)$, in Eq.(\ref{eqn:ResRate}), we obtain \cite{SuppMat}
\begin{align}\label{eqn:Res-Rate-a}
 & {\dot {\cal F}_{T}^{(a)}}(\Delta E) = \f{\Delta E}{2\pi} \biggl\{\f{1}{\e^{2\pi\Delta E/a}-1}
 + {\cal NT}\biggr\},
\end{align}
where
\begin{align}\label{eqn:NT}
 & {\cal NT} \equiv \l(\f{a}{2\pi\Delta E}\r) \biggl\{\cos(\Delta ET) \coth[(aT)/2] \nn \\
 &- \e^{-i\Delta ET}
 \times {}_{2}F_{1}\l[1,i(\Delta E/a);1+i(\Delta E/a);\e^{-aT}\r] \nn \\
 &- \e^{i\Delta ET}
 \times {}_{2}F_{1}\l[1,-i(\Delta E/a);1-i(\Delta E/a);\e^{-aT}\r]\biggr\},
\end{align}
with ${}_{2}F_{1}\l[\alpha,\beta;\gamma;z\r]$ denoting Gauss hypergeometric function
\cite{Abramowitz1972,Gradshteyn2007}.
Note that the second term in the above expression for finite-time accelerating detector response rate,
\ie, ${\cal NT}$, is nothing but the
transient effects that originate, as for the case of an inertial detector in Sec.\ref{Sec.-III},
from the finite-time interaction between the detector and
the (simple harmonic) field modes. Moreover, the first term in the above expression
is purely a thermal term, but the transient terms, ${\cal NT}$,
are non-thermal.
Therefore, due to the presence of transient effects in the finite-time
accelerating detector, the
detector cannot get thermalized with the Unruh bath, \ie,
\begin{align}
 \f{{\dot {\cal F}_{T}^{(a)}}(\Delta E)}{{\dot {\cal F}_{T}^{(a)}}(-\Delta E)}
 \neq \e^{-2\pi\Delta E/a},
\end{align}
for an arbitrary duration of time, $T$.
In other words, the presence of non-thermal terms, ${\cal NT}$,
obstructs the thermalization of finite-time uniformly accelerating detector
with the Unruh bath in its comoving frame.

For the finite-time accelerating detector response rate,
${\dot {\cal F}_{T}^{(a)}}(\Delta E)$, to be {\it purely}
thermal, the detector is in principle required to be interacting with the quantum field
for an {\it infinitely} long period of time, \ie, $T\to\infty$, so that
\begin{align}
 \lim_{T \to \infty} {\cal NT}=0.
\end{align}
However, in realistic setups that aim to test the existence of
Unruh effect, it is important to estimate the amount of time, say, $\tau_{\rm th}$, that is
required for the detector to interact with the field, so that the non-thermal contribution,
${\cal NT}$, would be {\it negligibly} small (see also \cite{Louko2016}), when compared to
the purely thermal piece in finite-time response rate of the accelerating detector in
Eq.(\ref{eqn:Res-Rate-a}), \ie,
\begin{align}
 \lim_{T \to \tau_{\rm th}} \f{{\dot {\cal F}_{T}^{(a)}}(\Delta E)}{{\dot {\cal F}_{T}^{(a)}}(-\Delta E)}
 \approx \e^{-2\pi\Delta E/a}.
\end{align}
We call the time duration, $\tau_{\rm th}$, {\it thermalization time}, since this is the time period that
is required for the detector in its comoving frame to get (almost) thermalized with the Unruh bath.

Note that the non-thermal transient terms, ${\cal NT}$, in the expression for
finite-time accelerating detector response rate in Eq.(\ref{eqn:Res-Rate-a})
is purely oscillatory with respect to the dimensionless variable, $\Delta ET$.
Therefore, in the limit $\Delta ET\gg1$,
the transient terms, ${\cal NT}$, can be averaged to zero, \ie,
\begin{align}
 \lim_{\Delta ET \to \infty} {\cal NT} = 0,
\end{align}
irrespective of the values of $(aT)$ and $(\Delta E/a)$ in the transient terms, ${\cal NT}$,
in Eq.(\ref{eqn:NT}). Hence,
the energy gap of the detector, $\Delta E$, sets the time scale, as for the case of inertial
detector discussed in Sec.(\ref{Sec.-III}), after which the non-thermal transient
terms, ${\cal NT}$, start becoming insignificant, \ie, $\tau_{\rm th} \gtrsim (\Delta E)^{-1}$.

In the forth coming section, we explicate the exact form of the thermalization-time,
$\tau_{\rm th}$, in the regimes of interest where the acceleration, $a$, of the detector to be
much less than the energy gap of the detector, $\Delta E$, \ie, $a\ll\Delta E$; and also
in the regime where $a\gg\Delta E$.
\section{How long to wait for thermality?}\label{Sec.-V}
Let us now turn to find the thermalization time -- the time duration,
$\tau_{\rm th}$, that a finite-time uniformly accelerating UD detector requires for it to get (almost)
thermalized with the Unruh bath in its comoving frame. For the Unruh bath in the comoving
frame of a uniformly accelerating detector to achieve even
a small Unruh temperature, say, 1K, the detector is required to move at
an extremely high accelerations that is of the order of $10^{21} m/s^2$
\cite{Takagi1986,Fulling1987,Wald1991,Crispino2008,Earman2011}. Due to this reason, the primary aim of any
promising experimental proposal that attempts to test Unruh effect is to enhance the physical
effects that originate due to the presence of Unruh bath
in the regime where the acceleration, $a$, of the detector
is sufficiently small, particularly with respect to the energy gap, $\Delta E$, of the detector,
\ie, $a\ll\Delta E$.
Therefore, in this section, we find the thermalization time, $\tau_{\rm th}$, of a detector that
is moving at small and large accelerations, specifically, in the regimes where $a\ll\Delta E$ and
$a\gg\Delta E$.
\subsection{In the limit $a \ll \Delta E$}
In the regime where the acceleration, $a$, of the detector is much smaller than the energy
gap of the detector, $\Delta E$, the expression for finite-time accelerating detector response rate,
${\dot {\cal F}_{T}^{(a)}}(\Delta E)$ in Eq.(\ref{eqn:Res-Rate-a}) gets reduced to \cite{SuppMat}
\begin{align}\label{eqn:Res-Rate-a-ll-E}
 & {\dot {\cal F}^{(a)}_{T}}(\Delta E) = \f{\Delta E}{2\pi}
 \biggl\{-\Theta(-\Delta E) + {\rm sgn}(\Delta E) \e^{-2\pi|\Delta E|/a} \nn \\
 &+ \f{\cos(\Delta ET)}{\pi\Delta ET}
 + \f{1}{\pi} {\rm Si}(\Delta ET) - \f{1}{2} {\rm sgn}(\Delta E) \nn \\
 &+ \l(\f{a}{\pi\Delta E}\r) \cos(\Delta ET) + \O\l[(a/\Delta E)^2\r]\biggr\}.
\end{align}
Few comments regarding the above expression are in order: Firstly,
the acceleration independent terms in the above expression matches exactly with the
expression for finite-time inertial detector response rate in Eq.(\ref{eqn:ResRateInertial}). Second,
the terms in the first line of the above expression correspond to purely thermal effects due to
the Unruh bath in the comoving frame of the detector, and the terms in the second and third lines
are due to the transient effects in the finite-time accelerating detector.
Third, for a sufficiently long period of time, $T$,
the oscillatory transient terms average out to be insignificantly small.
Therefore, 
for the purpose of finding the thermalization time, $\tau_{\rm th}$,
we define the averaged finite-time response rate,
${\dot {\widetilde {\cal F}}}^{(a)}_{T}(\Delta E)$,
of the accelerating detector as
\begin{align}\label{eqn:Avg-Res-Rate}
 {\dot {\widetilde {\cal F}}}^{(a)}_{T}(\Delta E) \equiv \f{1}{(T-T_0)}
 \int_{T_0}^{T} d\tau {\dot {\cal F}^{(a)}}_{\tau}(\Delta E),
\end{align}
where the finite-time response rate is averaged from an initial time,
$T_0$, to an arbitrary final time, $T$.

Making use of the expression for finite-time accelerating detector response rate in
Eq.(\ref{eqn:Res-Rate-a-ll-E}) in the definition for averaged finite-time
accelerating detector response rate in Eq.(\ref{eqn:Avg-Res-Rate}), we obtain
\begin{align}\label{eqn:Avg-Res-Rate-a-ll-E}
 {\dot {\widetilde {\cal F}}}^{(a)}_{T}(\Delta E) &\approx \f{\Delta E}{2\pi}
 \biggl\{-\Theta(-\Delta E) + {\rm sgn}(\Delta E) \e^{-2\pi|\Delta E|/a} \nn \\
 &+ \f{1}{\pi\Delta ET} \biggl[2\sin^2(\Delta ET/2) \nn \\
 &+ \Delta ET \l({\rm Si}(\Delta ET) - \f{\pi}{2} {\rm sgn}(\Delta E)\r) \nn \\
 &+ {\rm Ci}(\Delta ET)-{\rm Ci}(1)\biggr]\biggr\},
\end{align}
where we choose the initial time, $T_{0}$, to be of the order of $(\Delta E)^{-1}$, since $(\Delta E)^{-1}$
is the least time scale above which the expression for finite-time accelerating detector response in
Eq.(\ref{eqn:Res-Rate-a-ll-E}) remains valid, and
${\rm Ci}(x)$ is the cosine integral \cite{Abramowitz1972}.

\begin{figure}[!htb]
\begin{center}
\includegraphics[width=8 cm,height=5cm]{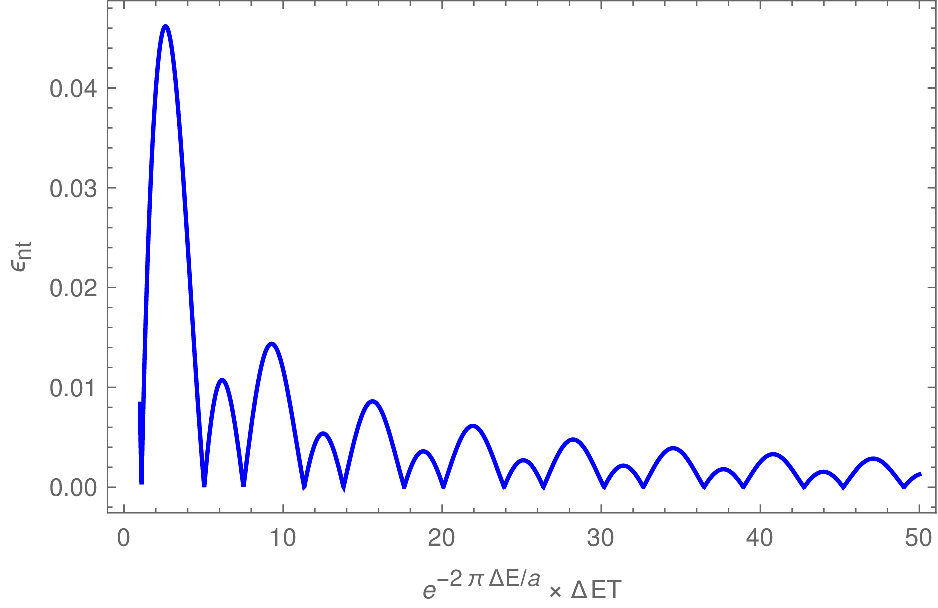} 
\caption{The non-thermal parameter, $\varepsilon_{\rm nt}$, is plotted with respect to the interaction
time, $T$, in the small acceleration limit, $a\ll\Delta E$. Since the response rate is exponentially
suppressed, \ie, $\sim \e^{-2\pi \Delta E/a}$ for the accelerations of the detector, $a\ll\Delta E$,
the interaction time, $T$, has to be exponentially large as $T \sim (\Delta E)^{-1} \e^{-2\pi \Delta E/a}$.}
\label{fig:allE}
\end{center}
\end{figure}

Furthermore, we define a parameter, $\varepsilon_{\rm nt}$, called non-thermal parameter, as
\begin{align}\label{eqn:nt-parameter}
 \varepsilon_{\rm nt} \equiv \l| \f{{\dot {\widetilde {\cal F}}}^{(a)}_{T}(\Delta E)}
 {{\dot {\widetilde {\cal F}}}^{(a)}_{\infty}(\Delta E)}-1 \r|,
\end{align}
that quantifies the presence of non-thermal transient effects in the finite-time accelerating
detector. An extremely small value of the non-thermal parameter, \ie,
$\varepsilon_{\rm nt}\ll1$, means a feeble presence of the non-thermal transient effects.
Substituting the expression for averaged finite-time response rate in
Eq.(\ref{eqn:Avg-Res-Rate-a-ll-E}) in the expression for the non-thermal parameter,
$\varepsilon_{\rm nt}$, in Eq.(\ref{eqn:nt-parameter}), we obtain
\begin{align}
 \varepsilon_{\rm nt} &\approx \f{\e^{2\pi|\Delta E|/a}}{|\Delta E|T}.
\end{align}
Therefore, demanding an arbitrary value for the non-thermal parameter,
$\varepsilon_{\rm nt}$, determines the corresponding value of the thermalization-time, $\tau_{\rm th}$,
and it is not surprising that lower the
non-thermality parameter, $\varepsilon_{\rm nt}$, higher would be the thermalization-time,
$\tau_{\rm th}$ (see Fig.(\ref{fig:allE})). Specifically,
if one demands a small value for the non-thermal parameter, $\varepsilon_{\rm nt}$, say,
$\delta$, \ie, $\varepsilon_{\rm nt}=\delta$, where $\delta\ll1$, then one can find the corresponding
thermalization-time, $\tau_{\rm th}$, as
\begin{align}\label{eqn:tau-th-a-ll-E}
  \tau_{\rm th} \sim \f{\e^{2\pi|\Delta E|/a}}{\Delta E\delta}.
\end{align}
Since the non-inertial contribution to the finite-time accelerating detector response rate,
${\dot {\cal F}^{(a)}_{T}}(\Delta E)$, is exponentially suppressed in the limit $a\ll\Delta E$,
we obtain the thermalization time, $\tau_{\rm th}$, to be exponentially large, \ie, of the order of
$\e^{2\pi|\Delta E|/a}$. Therefore, even if one could manage to perform a precision measurement
in measuring an extremely small response rate, \ie,
${\dot {\cal F}^{(a)}_{T}}(\Delta E) \propto \e^{-2\pi|\Delta E|/a}$ (though it is almost
impossible), the time that is required for the non-thermal transient terms to become insignificant,
and the detector to get (almost) thermalized with the Unruh bath is unrealistically large.
For example, suppose the detector is moving with a uniform acceleration, $a \sim 10^{9}m/s^2$;
the energy gap of the detector, $\Delta E$, is of the order of
1 MHz; and if one demands the non-thermal parameter, $\varepsilon_{\rm nt}$, to be of the
order of $10^{-6}$, then the thermalization time, $\tau_{\rm th}$, turns out to be
$\tau_{\rm th} \sim \e^{10^5}s$, which is much larger than the age of the universe! However,
this exponentially long waiting time can be reduced to a polynomial time when one tunes the switching
function, $\chi(\tau)$, such that the interaction between the UD detector and the quantum field is switched
on and off in a slow manner, and is long enough compared to the total duration of interaction \cite{Louko2016}.

It is evident from the above discussion that a detector moving at small accelerations,
\ie, $a\ll\Delta E$, in (3+1)-dimensional
Minkowski spacetime cannot achieve thermality with the Unruh bath within any realistic duration
of its proper time. Recent proposals
\cite{Lochan2020,Stargen2022,Navdeep2023,Stargen2024,Louko2024}
in literature predict a stark enhancement in the infinite-time response rate,
${\dot {\cal F}^{(a)}_{\infty}}(\Delta E)$, of detectors moving
at small accelerations, particularly, when boundary conditions are imposed on the field
modes.
Therefore,
it is clear from the expression for non-thermal parameter,
$\varepsilon_{\rm nt}$, in Eq.(\ref{eqn:nt-parameter}) that the thermalization time, $\tau_{\rm th}$,
for a detector to reach (near) thermality could be finite and viable when the detector is investigated in
cavity setups, and an attempt along this line of thought is currently in progress.
\subsection{In the limit $a \gg \Delta E$}
In the regime where the acceleration, $a$, of the detector is much larger than the energy
gap of the detector, $\Delta E$, the finite-time response rate,
${\dot {\cal F}_{T}^{(a)}}(\Delta E)$, of the accelerating detector in Eq.(\ref{eqn:Res-Rate-a})
gets reduced to
\begin{align}\label{eqn:Res-Rate-a-gg-E}
 {\dot {\cal F}_{T}^{(a)}}(\Delta E) &= \f{\Delta E}{2\pi} \biggl\{\f{a}{2\pi\Delta E}
 - \l(\f{a}{2\pi\Delta E}\r) \cos(\Delta ET) \nn \\
 &+ \O\l(\e^{-a/\Delta E}\r)\biggr\}.
\end{align}
Few comments regarding the above expression are in order: Firstly, the first term corresponds to
purely thermal effects due to the Unruh bath in the comoving frame of the detector.
Second, similar to the case of $a\ll\Delta E$,
the non-thermal transient terms are oscillatory, which get averaged out to be insignificant when
the detector is allowed to interact with the field for sufficiently long duration of time.
Third, both the thermal terms and the non-thermal transient terms grow in a similar fashion, \ie,
linear in $(a/\Delta E)$,
and due to which the thermalization time, $\tau_{\rm th}$, would be small when
compared to the same in the limit $a\ll\Delta E$. Specifically, in the expression for finite-time
accelerating detector response rate in the limit $a\ll\Delta E$
in Eq.(\ref{eqn:Res-Rate-a-ll-E}), the purely thermal term
is exponentially small, \ie, $\e^{-2\pi\Delta E/a}$, but the non-thermal transient terms
fall off with respect to time as $T^{-1}$. Therefore, the thermalization time,
$\tau_{\rm th}$, in the limit $a\ll\Delta E$ needs to be exponentially large, \ie, of the
order of $\e^{2\pi\Delta E/a}$, to make the non-thermal transient terms negligibly small, when
compared with the purely thermal term, so that the detector
gets (almost) thermalized with the Unruh bath.

\begin{figure}[!htb]
\begin{center}
\includegraphics[width=8 cm,height=5cm]{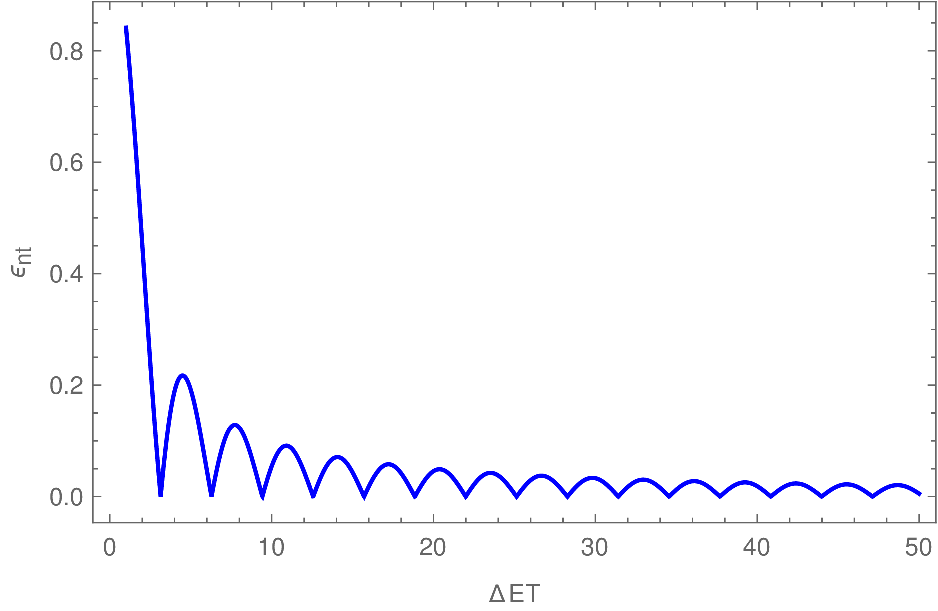} 
\caption{The non-thermal parameter, $\varepsilon_{\rm nt}$, is plotted with respect to the interaction
time, $T$, in the large acceleration limit, $a\gg\Delta E$. Unlike in the
low acceleration limit, $a\ll\Delta E$, the non-thermal parameter decreases substantially in a short
period of interaction time, $T$.}
\label{fig:aggE}
\end{center}
\end{figure}

To find the thermalization time, $\tau_{\rm th}$, in the limit $a\gg\Delta E$,
we substitute the expression for finite-time accelerating detector response rate,
${\dot {\cal F}_{T}^{(a)}}(\Delta E)$, in Eq.(\ref{eqn:Res-Rate-a-gg-E}) in the expression
for averaged finite-time detector response rate in Eq.(\ref{eqn:Avg-Res-Rate}), we obtain
\begin{align}
 {\dot {\widetilde {\cal F}}}^{(a)}_{T}(\Delta E) &\approx \f{\Delta E}{2\pi}
 \biggl\{\f{a}{2\pi\Delta E}
 - \l(\f{a}{2\pi\Delta E}\r) \f{\sin(\Delta ET)}{\Delta ET}\biggr\},
\end{align}
where we choose the initial time, $T_{0}$, to be of the order of $a^{-1}$, since $a^{-1}$
is the time scale above which the expression for finite-time accelerating detector response
in Eq.(\ref{eqn:Res-Rate-a-gg-E}) remains valid.
If one demands a small value, say, $\delta\ll1$, for the non-thermal parameter, $\varepsilon_{\rm nt}$,
\ie, $\varepsilon_{\rm nt}=\delta$, then we find the corresponding thermalization-time,
$\tau_{\rm th}$, to be (see Fig.(\ref{fig:aggE}))
\begin{align}
 \tau_{\rm th} \sim \f{1}{\Delta E\delta}.
\end{align}
Unlike the thermalization time, $\tau_{\rm th}$, in the limit $a\ll\Delta E$, as shown in the expression
in Eq.(\ref{eqn:tau-th-a-ll-E}), the thermalization time for large accelerations, $a\gg\Delta E$,
is independent of the acceleration, $a$, of the detector.

Though the thermalization time, $\tau_{\rm th}$, can be finite for large accelerations, \ie,
$a \gg \Delta E$, achieving large accelerations is challenging
\cite{Takagi1986,Fulling1987,Wald1991,Crispino2008,Earman2011}, due to which the proposals
to test Unruh effect aim at enhancing the effects of Unruh bath at sufficiently low accelerations
\cite{Scully2003,Lochan2020,Stargen2022,Navdeep2023,Stargen2024,Navdeep2022,
Majhi2024,ChenTajima1999,Habs2006,Habs2008,Matsas2001,Anglin2000,
Reznik2008,Louko2020,Mann2011,Ralph2022,Fuentes2010,Sudhir2022,Rogers1988,Kalinski2005}.
\section{Conclusion}\label{Sec.-VI}
We have investigated the response of a UD detector, that
is interacting with the massless scalar field for a finite duration of proper time, $T$, of the
detector. We explicitly evaluated the finite-time response rate of an inertial detector, and
a uniformly accelerating detector. We showed that the finite-time response rate of the accelerating
detector can be written as a sum of purely thermal terms, which are independent of time, $T$,
and non-thermal transient terms, which depend on time, $T$.
The non-thermal transient terms in the finite-time response rate of the
accelerating detector were shown to be oscillatory with respect to the variable, $(\Delta ET)$,
where $\Delta E$ is the energy gap of the detector.
Therefore, the oscillatory transient terms may get averaged out to be insignificant in the limit
$T\gg(\Delta E)^{-1}$, which implies the energy gap of the detector, $\Delta E$, sets a time scale
about which the transient effects fade off.

We calculated the finite-time accelerating detector response rate for small accelerations,
$a\ll\Delta E$, and also for large accelerations, $a\gg\Delta E$. We quantified the presence
of non-thermal effects in the finite-time detector response rate by introducing a (dimensionless)
parameter, $\varepsilon_{\rm th}$, called the non-thermal parameter. Specifically,
a lower value for the non-thermal parameter, $\varepsilon_{\rm th}$, signifies the detector
to be closer to thermality. We found, in both the small and large accelerations limit, that
smaller the non-thermal parameter, $\varepsilon_{\rm th}$, higher will be
the thermalization time, $\tau_{\rm th}$.
Specifically, in the low acceleration limit,
$a\ll\Delta E$, the thermalization time, $\tau_{\rm th}$, is exponentially large, ie,
$\sim (\Delta E)^{-1} \times \e^{2\pi|\Delta E|/a}/\varepsilon_{\rm th}$; and in the large
acceleration limit, $a\gg\Delta E$, the thermalization time is of the order of
$(\Delta E)^{-1}/\varepsilon_{\rm th}$.

We note that the thermalization time, $\tau_{\rm th}$, at low accelerations, $a\ll\Delta E$,
being exponentially large
is due to the fact that the purely thermal effects in the finite-time accelerating detector
response rate falls off exponentially as $\sim \e^{-2\pi|\Delta E|/a}$.
Since the non-thermal effects in finite-time accelerating detector response fall off as
$1/T$, the detector has to interact with the Unruh bath for exponentially long period
of time, so that the non-thermal transient terms become insignificant compared to the
thermal effects. Therefore, in this regard, if the thermal effects in the finite-time accelerating detector
are made to be substantially enhanced, when compared to $\e^{-2\pi\Delta E/a}$, then
the thermalization time, $\tau_{\rm th}$, could be brought down to more realistic values.
Moreover, recent proposals \cite{Lochan2020,Stargen2022,Navdeep2023,Stargen2024,Louko2024}
in literature suggest physical scenarios
where boundary conditions are imposed on field modes, such that a sharp increase in
the density of field modes could lead to an exponential enhancement of
the detector response rate at appreciably small accelerations. Therefore, following this line of thought
could make the thermalization time, $\tau_{\rm th}$, in the limit, $a \ll \Delta E$, finite
and viable, and this attempt is currently in progress.
\section{Acknowledgements}
I would like to thank Ross Jenkinson, Jorma Louko, and Vivishek Sudhir for interesting and helpful
comments on the manuscript.
\begin{widetext}
\section{Supplemental Material: Calculation details of Eq.(7)}
The finite-time response, ${\cal F}_{T}(\Delta E)$, of a UD detector that is moving on a given
spacetime trajectory, say, ${\tilde x}(\tau)$, is defined as
\begin{align}
 {\cal F}_{T}(\Delta E) &\equiv \int_{-T/2}^{T/2} d\tau \int_{-T/2}^{T/2} d\tau'
 e^{-i\Delta E(\tau-\tau')} {\cal W}(\tau,\tau'),
\end{align}
in which we make a variable change $u=\tau-\tau'$, $v=(\tau+\tau')/2$ to obtain
\begin{align}
 {\cal F}_{T}(\Delta E) &= \int_{-T}^{0} du \int_{-(u+T)/2}^{(u+T)/2} dv~ e^{-i\Delta Eu} 
 {\cal W}\l(v+u/2,v-u/2\r) \nn \\ \nn \\
 &+ \int_{0}^{T} du \int_{(u-T)/2}^{(-u+T)/2} dv~ e^{-i\Delta Eu}
 {\cal W}\l(v+u/2,v-u/2\r). 
\end{align}
If the detector moves along the integral curve of a Killing vector field corresponding to the
background Minkowski spacetime, then the Wightman function, ${\cal W}(\tau,\tau')$, becomes a function
of $(\tau-\tau)$, \ie, ${\cal W}(\tau-\tau',0)$ \cite{Takagi1986,Fulling1987,Wald1991,Crispino2008,Earman2011}.
Therefore, the expression for the finite-time detector response, ${\cal F}_{T}(\Delta E)$, becomes
\begin{align}
 {\cal F}_{T}(\Delta E) &= \int_{-T}^{0} du \int_{-(u+T)/2}^{(u+T)/2} dv~ e^{-i\Delta Eu} 
 {\cal W}\l(u,0\r)
 + \int_{0}^{T} du \int_{(u-T)/2}^{(-u+T)/2} dv~ e^{-i\Delta Eu}
 {\cal W}\l(u,0\r),
\end{align}
which can be simplified to
\begin{align}\label{eqn:App-Response}
 {\cal F}_{T}(\Delta E) &= T\int_{-T}^{T} du e^{-i\Delta Eu} {\cal W}(u,0)
 - \int_{-T}^{T} du~{\rm sgn}(u)~u~e^{-i\Delta Eu} {\cal W}(u,0).
\end{align}

Defining the finite-time response rate, ${\dot {\cal F}}_{T}(\Delta E)$, of the detector as
\begin{align}
 {\dot {\cal F}}_{T}(\Delta E) \equiv \f{d{\cal F}(\Delta E)}{d\tau},
\end{align}
and making use of Leibniz's theorem for differentiation of integrals \cite{Abramowitz1972},
in the expression for finite-time detector response in Eq.(\ref{eqn:App-Response}), we obtain the
expression for finite-time detector response rate, ${\dot {\cal F}}_{T}(\Delta E)$, as
\begin{align}
 {\dot {\cal F}}_{T}(\Delta E) = \int_{-T}^{T} du~ e^{-i\Delta Eu} {\cal W}(u,0).
\end{align}
\section{Calculation details of Eq.(12)}
Substituting the trajectory information of the uniformly accelerating detector, i.e.,
$(a^{-1}{\rm sinh}a\tau,\rho_0,\theta_0,a^{-1}{\rm cosh}a\tau)$, in the expression for Wightman
function,
\begin{align}
 {\cal W}(x,x') &= \f{1}{(2\pi)^2} \sum_{m=-\infty}^{\infty}
 \int_0^\infty \d q~ q J_{m}(q\rho) J_{m}(q\rho')
 \int_{-\infty}^{\infty} \f{\d k_z}{2\omega_k}
 \e^{-i\omega_k (t-t')} \e^{im(\theta-\theta')} \e^{ik_z(z-z')},
\end{align}
we obtain
\begin{align}
 {\cal W}(\tau,\tau') = \f{1}{(2\pi)^2} \int_0^\infty \d q~ q
 \int_{-\infty}^{\infty} \f{\d k_z}{2\omega_k}
 {\rm exp}\l\{-i\f{\omega_k}{a} ({\rm sinh}a\tau-{\rm sinh}a\tau')
 +i\f{k_z}{a} \l({\rm cosh}a\tau-{\rm cosh}a\tau'\r)\r\},
\end{align}
in which making a variable change, $u=\tau-\tau'$, and $v=(\tau+\tau')/2$, leads to
\begin{align}
 {\cal W}(\tau,\tau') = \f{1}{(2\pi)^2} \int_0^\infty \d q~q
 \int_{-\infty}^{\infty} \f{\d k_z}{2\omega_k}
 {\rm exp}\l\{-\f{2i}{a}{\rm sinh}(au/2)
 \l(\omega_k {\rm cosh}av-k_z {\rm sinh}av\r)\r\}.
\end{align}
Making a further variable change as
\begin{align}
 \omega_k'=\omega_k {\rm cosh}av-k_z {\rm sinh}av,~~
 k_z'=k_z {\rm cosh}av-\omega_k {\rm sinh}av,
\end{align}
one obtains
\begin{align}
  {\cal W}(u,0) = \f{1}{(2\pi)^2} \int_0^\infty \d q~q
 \int_{-\infty}^{\infty} \f{\d k_z'}{2\omega_k'} e^{-i\Delta Eu}
 \e^{-2i(\omega_k'/a) {\rm sinh}(au/2)}.
\end{align}
Substituting the above expression for Wightman function for the accelerating detector
in the expression for transition probability rate
\begin{align}
 {\cal {\dot F}}^{(a)}_{T}(\Delta E) = \int_{-T}^{T} \d u \e^{-i\Delta E u} {\cal W}(u,0),
\end{align}
we obtain
\begin{align}
 {\cal {\dot F}}^{(a)}_{T}(\Delta E) = \f{T}{(2\pi)^2} \int_0^\infty \d q~ q
 \int_{-\infty}^{\infty} \f{\d k_z}{2\omega_k}
 \int_{-1}^{1} \d {\tilde u}~e^{-i\Delta ET {\tilde u}}~
 \e^{-2i(\omega_k/a) {\rm sinh}(aT{\tilde u}/2)}.
\end{align}
which can further be rewritten as
\begin{align}
 {\cal {\dot F}}^{(a)}_{T}(\Delta E) &= \f{T}{(2\pi)^2} \int_0^\infty \d q~ q
 \int_{-\infty}^{\infty} \f{\d k_z}{2\omega_k} \nn \\
 &\times \biggl\{\int_{-\infty}^{\infty} \d {\tilde u} e^{-i\Delta ET {\tilde u}}
 {\rm exp}\l\{-2i(\omega_k/a) {\rm sinh}(aT{\tilde u}/2)\r\} \nn \\
 &- \int_{-\infty}^{\infty} \d {\tilde u} \Theta({\tilde u}-1) \e^{-i\Delta ET {\tilde u}}
 {\rm exp}\l\{-2i(\omega_k/a) {\rm sinh}(aT{\tilde u}/2)\r\} \nn \\
 &- \int_{-\infty}^{\infty} \d {\tilde u} \Theta(-{\tilde u}-1) \e^{-i\Delta ET {\tilde u}}
 {\rm exp}\l\{-2i(\omega_k/a) {\rm sinh}(aT{\tilde u}/2)\r\}\biggr\},
\end{align}
where $\Theta(x)$ is known as Heaviside theta function \cite{Gradshteyn2007}. Making use of the
integral representation of Heaviside theta function \cite{Gradshteyn2007}
\begin{align}
 \Theta(x) = \f{1}{2\pi i} \int_{-\infty}^{\infty}
 \f{\d\alpha}{\alpha-i\varepsilon} \e^{i\alpha x},
\end{align}
we obtain
\begin{align}
 {\dot {\cal F}}^{(a)}_{T}(\Delta E) &= \f{T}{(2\pi)^2} \int_0^\infty \d q~ q
 \int_{-\infty}^{\infty} \f{\d k_z}{2\omega_k}
 \biggl\{\int_{-\infty}^{\infty} \d {\tilde u} e^{-i\Delta ET {\tilde u}}
 {\rm exp}\l\{-2i(\omega_k/a) {\rm sinh}(aT{\tilde u}/2)\r\} \nn \\
 &+ \f{\e^{-i\Delta ET}}{2\pi i} \int_{-\infty}^{\infty}
 \f{\d\alpha}{\alpha-\Delta ET+i\varepsilon} \e^{i\alpha}
 \times \int_{-\infty}^{\infty} \d {\tilde u} \e^{-i\alpha{\tilde u}}
 {\rm exp}\l\{-2i(\omega_k/a) {\rm sinh}(aT{\tilde u}/2)\r\} \nn \\
 &- \f{\e^{i\Delta ET}}{2\pi i} \int_{-\infty}^{\infty}
 \f{\d\alpha}{\alpha-\Delta ET-i\varepsilon} \e^{-i\alpha}
 \times \int_{-\infty}^{\infty} \d {\tilde u} \e^{-i\alpha{\tilde u}}
 {\rm exp}\l\{-2i(\omega_k/a) {\rm sinh}(aT{\tilde u}/2)\r\}\biggr\}.
\end{align}
Evaluating the ${\tilde u}$ integral leads to
\begin{align}
 {\dot {\cal F}}^{(a)}_{T}(\Delta E) &= \f{T}{(2\pi)^2} \l(\f{2}{aT}\r) \int_0^\infty \d q~q
 \biggl\{\e^{-\pi\Delta E/a} K^2_{i\Delta E/a}(q/a) \nn \\
 &+ \f{\e^{-i\Delta ET}}{2\pi i} \int_{-\infty}^{\infty}
 \f{\d\alpha}{\alpha-\Delta ET+i\varepsilon} \e^{i\alpha}
 \times \e^{-\pi\alpha/aT} K^2_{i\alpha/aT}(q/a) \nn \\
 &- \f{\e^{i\Delta ET}}{2\pi i} \int_{-\infty}^{\infty}
 \f{\d\alpha}{\alpha-\Delta ET-i\varepsilon} \e^{-i\alpha}
 \times \e^{-\pi\alpha/aT} K^2_{i\alpha/aT}(q/a)\biggr\},
\end{align}
and evaluating the $q$ integral further, we get
\begin{align}
 {\dot {\cal F}}^{(a)}_{T}(\Delta E) &= \f{a}{(2\pi)^2} \biggl\{2\pi (\Delta E/a)
 \f{1}{\e^{2\pi\Delta E/a}-1} \nn \\
 &+ \f{\e^{-i\Delta ET}}{2\pi i} \int_{-\infty}^{\infty}
 \f{\d\alpha}{\alpha-\Delta ET+i\varepsilon} \e^{i\alpha}
 \times \e^{-\pi\alpha/aT} \times (\pi\alpha/aT) {\rm csch}(\pi\alpha/aT) \nn \\
 &- \f{\e^{i\Delta ET}}{2\pi i} \int_{-\infty}^{\infty}
 \f{\d\alpha}{\alpha-\Delta ET-i\varepsilon} \e^{-i\alpha}
 \times \e^{-\pi\alpha/aT} \times (\pi\alpha/aT) {\rm csch}(\pi\alpha/aT)\biggr\}.
\end{align}
Making use of the series representation of ${\rm csch}(\pi z)$
\begin{align}
 {\rm csch}(\pi z) = \f{1}{\pi z} + \f{2z}{\pi} \sum_{n=1}^{\infty} \f{(-1)^n}{z^2+n^2},
\end{align}
we evaluate the $\alpha$ integral as
\begin{align}
 {\dot {\cal F}}^{(a)}_{T}(\Delta E) &= \f{\Delta E}{2\pi} \biggl\{\f{1}{\e^{2\pi\Delta E/a}-1}
 + \l(\f{a}{\pi\Delta E}\r) \Re\biggl[\e^{-i\Delta ET}
 \sum_{n=1}^{\infty} \f{n}{n+i(\Delta E/a)+\varepsilon} \e^{-n(aT)}\biggr]\biggr\},
\end{align}
where the series can be rewritten as an integral as
\begin{align}
 {\dot {\cal F}}^{(a)}_{T}(\Delta E) &= \f{\Delta E}{2\pi} \biggl\{\f{1}{\e^{2\pi\Delta E/a}-1}
 + \l(\f{a}{\pi\Delta E}\r) \times \f{(aT)}{4} \Re\biggl[\int_{1}^{\infty} \d x'
 \f{\e^{-i(\Delta ET-i\varepsilon) x'}}{\sinh^2[(aT)x'/2]}\biggr]\biggr\}.
\end{align}

Performing an integration by parts in the $x'$ integral, we obtain
\begin{align}
 {\dot {\cal F}}^{(a)}_{T}(\Delta E) &= \f{\Delta E}{2\pi} \biggl\{\f{1}{\e^{2\pi\Delta E/a}-1} \nn \\
 &+ \l(\f{a}{2\pi\Delta E}\r) \biggl(\cos(\Delta ET) \coth[(aT)/2] + (\Delta ET)
 \Im\l[\int_{1}^{\infty} \d x' \e^{-i(\Delta ET-i\varepsilon) x'} \coth[(aT)x'/2]\r]\biggr)\biggr\},
\end{align}
and making use of the series representation
\begin{align}
 \coth z = \f{1}{z} + \sum_{n=1}^{\infty} \l(\f{1}{z+in\pi} + \f{1}{z-in\pi}\r),
\end{align}
we obtain
\begin{align}
 {\dot {\cal F}}^{(a)}_{T}(\Delta E) &= \f{\Delta E}{2\pi} \Biggl\{\f{1}{\e^{2\pi\Delta E/a}-1}
 + \f{1}{\pi} {\rm Si}(\Delta ET)-\f{1}{2}{\rm sgn}(\Delta E) \nn \\
 &+ \l(\f{a}{2\pi\Delta E}\r) \Biggl[\cos(\Delta ET) \coth[(aT)/2] \nn \\
 &+ (\Delta ET) \Im\l(\sum_{n=1}^{\infty} \int_{-\infty}^{\infty} \d x' \Theta(x'-1)
 \e^{-i(\Delta ET-i\varepsilon) x'}
 \l[\f{1}{(aT/2)x'+in\pi} + \f{1}{(aT/2)x'-in\pi}\r]\r)\Biggr]\Biggr\}.
\end{align}
Making use of the integral representation for Heaviside theta function \cite{Gradshteyn2007},
and performing the $x'$ integral, we obtain
\begin{align}
 & {\dot {\cal F}}^{(a)}_{T}(\Delta E) = \f{\Delta E}{2\pi} \Biggl\{\f{1}{\e^{2\pi\Delta E/a}-1}
 + \f{1}{\pi} {\rm Si}(\Delta ET)-\f{1}{2}{\rm sgn}(\Delta E)
 + \l(\f{a}{2\pi\Delta E}\r) \Biggl[\cos(\Delta ET) \coth[(aT)/2] \nn \\
 &+ \f{(\Delta ET)}{(aT)} \Im\biggl(\int_{-\infty}^{\infty}
 \f{\d\alpha}{\alpha-i\varepsilon} \e^{-i\alpha}
 \coth\l(\f{\pi}{aT} (\alpha-\Delta ET+i\varepsilon)\r)
 - \int_{-\infty}^{\infty} \f{\d\alpha}{\alpha-i\varepsilon}
 \e^{-i\alpha} {\rm sgn}(\alpha-\Delta ET)\biggr)\Biggr]\Biggr\}.
\end{align}
Evaluating the $\alpha$ integral, we obtain
\begin{align}
 {\dot {\cal F}}^{(a)}_{T}(\Delta E) &= \f{\Delta E}{2\pi} \Biggl\{\f{1}{\e^{2\pi\Delta E/a}-1} \nn \\
 &+ \l(\f{a}{2\pi\Delta E}\r) \Biggl[\cos(\Delta ET) \coth[(aT)/2]
 + (\Delta ET) \Im\biggl(2\e^{-i\Delta ET} \sum_{n=0}^{\infty}
 \f{\e^{-n(aT)}}{n(aT)+i\Delta ET}\biggr)\Biggr]\Biggr\},
\end{align}
and rewriting the series over $n$ as
\begin{align}
 {\dot {\cal F}}^{(a)}_{T}(\Delta E) &= \f{\Delta E}{2\pi} \Biggl\{\f{1}{\e^{2\pi\Delta E/a}-1} \nn \\
 &+ \l(\f{a}{2\pi\Delta E}\r) \Biggl[\cos(\Delta ET) \coth[(aT)/2]
 + 2(\Delta ET) \Im\biggl(\int_{1}^{\infty} dx
 \f{\e^{-i\Delta ETx}}{1-\e^{-(aT)x}}\biggr)\Biggr]\Biggr\}.
\end{align}
Evaluating the $x$ integral, we obtain
\begin{align}
 {\dot {\cal F}}^{(a)}_{T}(\Delta E) &= \f{\Delta E}{2\pi} \Biggl\{\f{1}{\e^{2\pi\Delta E/a}-1}
 + \l(\f{a}{2\pi\Delta E}\r) \biggl\{\cos(\Delta ET) \coth[(aT)/2] \nn \\
 &- \e^{-i\Delta ET}
 \times {}_{2}F_{1}\l[1,i(\Delta E/a);1+i(\Delta E/a);\e^{-aT}\r] \nn \\
 &- \e^{i\Delta ET}
 \times {}_{2}F_{1}\l[1,-i(\Delta E/a);1-i(\Delta E/a);\e^{-aT}\r]\biggr\}\Biggr\},
\end{align}
where ${}_{2}F_{1}\l[\alpha,\beta;\gamma;z\r]$ denotes Gauss hypergeometric function
\cite{Abramowitz1972,Gradshteyn2007}.
\section{Calculation details of Eq.(18)}
Since the Gauss hypergeometric function, ${}_{2}F_{1}[1,\beta;\beta+1;z]$, has the integral
representation as
\begin{align}
 {}_{2}F_{1}[1,\beta;\beta+1;z]=\beta \int_{0}^{1} dt~t^{\beta-1} \f{1}{(1-zt)},
\end{align}
we have
\begin{align}
 {}_{2}F_{1}\l[1,i(\Delta E/a);1+i(\Delta E/a);\e^{-aT}\r]
 = i(\Delta E/a) \e^{i\Delta ET} \int_{0}^{\e^{-aT}} dt~\f{t^{i(\Delta E/a)-1}}{\l(1-t\r)}.
\end{align}
Substituting $t=\e^{-(a/\Delta E)y}$, the above expression takes the form
\begin{align}
 {}_{2}F_{1}\l[1,i(\Delta E/a);1+i(\Delta E/a);\e^{-aT}\r]
 = i\e^{i\Delta ET}
 \int_{\Delta ET}^{\infty} \d y~\f{\e^{-iy}}{\l(1-\e^{-(a/\Delta E)y}\r)},
\end{align}
and employing this in the expression for finite-time acceleration detector response rate,
\begin{align}\label{eqn:Res-Rate-a}
 & {\dot {\cal F}_{T}^{(a)}}(\Delta E) = \f{\Delta E}{2\pi} \biggl\{\f{1}{\e^{2\pi\Delta E/a}-1}
 + {\cal NT}\biggr\},
\end{align}
where
\begin{align}\label{eqn:NT}
 & {\cal NT} \equiv \l(\f{a}{2\pi\Delta E}\r) \biggl\{\cos(\Delta ET) \coth[(aT)/2] \nn \\
 &- \e^{-i\Delta ET}
 \times {}_{2}F_{1}\l[1,i(\Delta E/a);1+i(\Delta E/a);\e^{-aT}\r] \nn \\
 &- \e^{i\Delta ET}
 \times {}_{2}F_{1}\l[1,-i(\Delta E/a);1-i(\Delta E/a);\e^{-aT}\r]\biggr\},
\end{align}
we obtain
\begin{align}
 {\dot {\cal F}}^{(a)}_{T}(\Delta E) &= \f{\Delta E}{2\pi} \biggl\{\f{1}{\e^{2\pi\Delta E/a}-1}
 + \l(\f{a}{2\pi\Delta E}\r) \cos(\Delta ET) \coth[(aT)/2] \nn \\
 &- \l(\f{a}{\pi\Delta E}\r) \int_{\Delta ET}^{\infty} \d y
 \f{\sin y}{\l(1-\e^{-(a/\Delta E)y}\r)}\biggr\}.
\end{align}
In the limit $a \ll \Delta E$, the above expression gets reduced to
\begin{align}
 {\dot {\cal F}}^{(a)}_{T}(\Delta E) &= \f{\Delta E}{2\pi} \biggl\{\f{1}{\e^{2\pi\Delta E/a}-1}
 + \l(\f{a}{2\pi\Delta E}\r) \cos(\Delta ET) \coth[(aT)/2] \nn \\
 &+ \f{1}{\pi} \l[{\rm Si}(\Delta ET) - \f{\pi}{2} {\rm sgn}(\Delta E)
 + (a/\Delta E) \cos (\Delta ET) + O\l[(a/\Delta E)^2\r]\r]\biggr\}.
\end{align}
\end{widetext}


\begin{thebibliography}{50}
\bibitem{Fulling1973}
Fulling, S. A.~Nonuniqueness of canonical field quantization in Riemannian space-time,
\href{https://doi.org/10.1103/PhysRevD.7.2850}{{\it Phys. Rev. D}, {\bf 7}:2850--2862, (1973)}.
%
\bibitem{Davies1975}
Davies, P. C. W. Scalar particle production in Schwarzschild and Rindler metrics,
\href{https://iopscience.iop.org/article/10.1088/0305-4470/8/4/022}{{\it J. Phys. A},
{\bf 8}:609--616, (1975)}.
%
\bibitem{Unruh1976}
Unruh, W. G. Notes on black-hole evaporation,
\href{https://doi.org/10.1103/PhysRevD.14.870}{{\it Phys. Rev. D},
{\bf 40}:870--892, (1976)}.
%
\bibitem{Takagi1986}
Takagi, S. Vacuum noise and stress induced by uniform acceleration—Hawking-Unruh effect in
Rindler manifold of arbitrary dimension,
\href{https://academic.oup.com/ptps/article/doi/10.1143/PTP.88.1/1938595}{{\it Prog. Theor. Phys. Suppl.},
{\bf 88}:1--142, (1986)}.
%
\bibitem{Fulling1987}
Fulling, S. A. and Ruijsenaars, S. N. M., Temperature, periodicity and horizons,
\href{https://www.sciencedirect.com/science/article/abs/pii/0370157387901360}{{\it Phys. Rep.},
{\bf 152}:135--176, (1987)}.
%
\bibitem{Wald1991}
Kay, B. S. and Wald, R. M., Theorems on the uniqueness and thermal properties of stationary,
nonsingular, quasifree states on spacetimes with a bifurcate Killing horizon,
\href{https://www.sciencedirect.com/science/article/abs/pii/037015739190015E}{{\it Phys. Rep.},
{\bf 207}:49--136, (1991)}.
%
\bibitem{Crispino2008}
Crispino, L. C. B., Higuchi, A., and Matsas, G. E. A., The Unruh effect and its applications,
\href{https://journals.aps.org/rmp/abstract/10.1103/RevModPhys.80.787}{{\it Rev. Mod. Phys.},
{\bf 80}:787--838, (2008)}.
%
\bibitem{Earman2011}
Earman, J., The Unruh effect for philosophers,
\href{https://www.sciencedirect.com/science/article/abs/pii/S1355219811000207}
{{\it Stud. Hist. Philos. Mod. Phys.}, {\bf 42}:81--97, (2011)}.
%
\bibitem{DeWitt1979}
DeWitt, B., in {\it General Relativity: An Einstein Centenary Survey}, edited by S. W. Hawking and W. Israel,
(Cambridge University Press, Cambridge, England, 1979).
%
\bibitem{Wald1984}
Unruh, W. G. and Wald, R. M., What happens when an accelerating observer detects a Rindler particle,
\href{https://journals.aps.org/prd/abstract/10.1103/PhysRevD.29.1047}
{{\it Phys. Rev. D}, {\bf 29}:1047--1056, (1984)}.
%
\bibitem{Merkli2006}
De Bievre, S. and Merkli, M., The Unruh effect revisited,
\href{https://iopscience.iop.org/article/10.1088/0264-9381/23/22/026}
{{\it Class. Quantum Grav.}, {\bf 23}:6525--6541, (2006)}.
%
\bibitem{Garay2021}
Arrechea, J., Barcelo, C., Garay, L. J., and Garcia-Moreno, G.,
Inversion of statistics and thermalization in the Unruh effect,
\href{https://journals.aps.org/prd/abstract/10.1103/PhysRevD.104.065004}
{{\it Phys. Rev. D}, {\bf 104}:065004, (2021)}.
%
\bibitem{Scully2003}
Scully, M. O., Kocharovsky, V. V., Belyanin, A., Fry, E., and Capasso, F.,
Enhancing Acceleration Radiation from Ground-State Atoms via Cavity Quantum Electrodynamics,
\href{https://journals.aps.org/prl/abstract/10.1103/PhysRevLett.91.243004}
{{\it Phys. Rev. Lett.}, {\bf 91}:243004, (2003)}.
%
\bibitem{Lochan2020}
Lochan, K., Ulbricht, H., Vinante, A., Goyal, S. K., Fry, E., and Capasso, F.,
Detecting Acceleration-Enhanced Vacuum Fluctuations with Atoms Inside a Cavity,
\href{https://journals.aps.org/prl/abstract/10.1103/PhysRevLett.125.241301}
{{\it Phys. Rev. Lett.}, {\bf 125}:241301, (2020)}.
%
\bibitem{Stargen2022}
Stargen, D. J. and Lochan, K., Cavity optimization for Unruh effect at small accelerations,
\href{https://journals.aps.org/prl/abstract/10.1103/PhysRevLett.129.111303}
{{\it Phys. Rev. Lett.}, {\bf 129}:111303, (2022)}.
%
\bibitem{Navdeep2023}
Arya, N. and Goyal, S. K., Lamb shift as a witness for quantum noninertial effects,
\href{https://doi.org/10.1103/PhysRevD.108.085011}
{{\it Phys. Rev. D}, {\bf 108}:085011, (2023)}.
%
\bibitem{Stargen2024}
Arya, N., Stargen, D. J., Lochan, K., and Goyal, S. K., Strong noninertial radiative shifts
in atomic spectra at low Accelerations,
\href{https://doi.org/10.1103/PhysRevD.110.085007}
{{\it Phys. Rev. D}, {\bf 110}:085007, (2024)}.
%
\bibitem{Navdeep2022}
Arya, N., Mittal, V., Lochan, K., and Goyal, S. K., Geometric phase assisted enhancement
of non-inertial cavity-QED effects,
\href{https://doi.org/10.1103/PhysRevD.106.045011}
{{\it Phys. Rev. D}, {\bf 106}:045011, (2022)}.
%
\bibitem{Majhi2024}
Barman, D., Ghosh, D., and Majhi, B. R., Mirror-enhanced acceleration induced
geometric phase: towards detection of Unruh effect,
\href{https://arxiv.org/abs/2405.07711}{arXiv:2405.07711 (2024)}.
%
\bibitem{ChenTajima1999}
Chen, P. and Tajima, T., Testing Unruh Radiation with Ultraintense Lasers,
\href{https://journals.aps.org/prl/abstract/10.1103/PhysRevLett.83.256}
{{\it Phys. Rev. Lett.}, {\bf 83}:256, (1999)}.
%
\bibitem{Habs2006}
Schutzhold, R., Schaller, G., and Habs, D., Signatures of the Unruh effect
from electrons accelerated by ultrastrong laser fields,
\href{https://journals.aps.org/prl/abstract/10.1103/PhysRevLett.97.121302}
{{\it Phys. Rev. Lett.}, {\bf 97}:121302, (2006)}.
%
\bibitem{Habs2008}
Schutzhold, R., Schaller, G., and Habs, D., Tabletop Creation of Entangled
Multi-keV Photon Pairs and the Unruh Effect,
\href{https://journals.aps.org/prl/abstract/10.1103/PhysRevLett.100.091301}
{{\it Phys. Rev. Lett.}, {\bf 100}:091301, (2008)}.
%
\bibitem{Rogers1988}
Rogers, J., Detector for the Temperaturelike Effect of Acceleration,
\href{https://journals.aps.org/prl/abstract/10.1103/PhysRevLett.61.2113}
{{\it Phys. Rev. Lett.}, {\bf 61}:2113, (1988)}.
%
\bibitem{Matsas2001}
Vanzella, D. A. T. and Matsas, G. E. A., Decay of Accelerated Protons and
the Existence of the Fulling-Davies-Unruh Effect,
\href{https://journals.aps.org/prl/abstract/10.1103/PhysRevLett.87.151301}
{{\it Phys. Rev. Lett.}, {\bf 87}:151301, (2001)}.
%
\bibitem{Anglin2000}
Garay, L. J. and Anglin, J. R., Sonic Analog of Gravitational Black
Holes in Bose-Einstein Condensates,
\href{https://journals.aps.org/prl/abstract/10.1103/PhysRevLett.85.4643}
{{\it Phys. Rev. Lett.}, {\bf 85}:4643, (2000)}.
%
\bibitem{Reznik2008}
Retzker, A., Cirac, J. I., Plenio, M. B., and Reznik, B., Methods for
Detecting Acceleration Radiation in a Bose-Einstein Condensate,
\href{https://journals.aps.org/prl/abstract/10.1103/PhysRevLett.101.110402}
{{\it Phys. Rev. Lett.}, {\bf 101}:110402, (2008)}.
%
\bibitem{Louko2020}
Gooding, C., Biermann, S., Erne, S., Louko, J., Unruh, W. G., Schmiedmayer, J.,
and Weinfurtner, S., Interferometric Unruh Detectors for Bose-Einstein Condensates,
\href{https://journals.aps.org/prl/abstract/10.1103/PhysRevLett.125.213603}
{{\it Phys. Rev. Lett.}, {\bf 125}:213603, (2020)}.
%
\bibitem{Mann2011}
Martin-Martinez, E., Fuentes, I., and Mann, R. B., Using Berry’s Phase to Detect
the Unruh Effect at Lower Accelerations,
\href{https://journals.aps.org/prl/abstract/10.1103/PhysRevLett.107.131301}
{{\it Phys. Rev. Lett.}, {\bf 107}:131301, (2011)}.
%
\bibitem{Ralph2022}
Quach, J. Q., Ralph, T. C., and Munro, W. J., Berry Phase from the Entanglement
of Future and Past Light Cones: Detecting the Timelike Unruh Effect,
\href{https://journals.aps.org/prl/abstract/10.1103/PhysRevLett.129.160401}
{{\it Phys. Rev. Lett.}, {\bf 129}:160401, (2022)}.
%
\bibitem{Kalinski2005}
Kalinski, M., Hawking radiation from trojan states in muonic hydrogen in strong
laser field. {\it Laser Physics}, {\bf 15}:10 (2005).
%
\bibitem{Fuentes2010}
Aspachs, M., Adesso, G., and Fuentes, I., Optimal Quantum Estimation of the
Unruh-Hawking Effect,
\href{https://journals.aps.org/prl/abstract/10.1103/PhysRevLett.105.151301}
{{\it Phys. Rev. Lett.}, {\bf 105}:151301, (2010)}.
%
\bibitem{Sudhir2022}
Soda, B., Sudhir, V., and Kempf, A., Acceleration-Induced Effects in
Stimulated Light-Matter Interactions,
\href{https://journals.aps.org/prl/abstract/10.1103/PhysRevLett.128.163603}
{{\it Phys. Rev. Lett.}, {\bf 128}:163603, (2022)}.
%
\bibitem{Svaiter1992}
Svaiter, B. F., and Svaiter, N. F., Inertial and noninertial particle detectors and vacuum fluctuations,
\href{https://journals.aps.org/prd/abstract/10.1103/PhysRevD.46.5267}
{{\it Phys. Rev. D}, {\bf 46}:5267, (1992)}.
%
\bibitem{Peres1993}
Higuchi, A., Matsas, G. E. A., and Peres, C. B., Uniformly accelerated finite-time detectors,
\href{https://journals.aps.org/prd/abstract/10.1103/PhysRevD.48.3731}
{{\it Phys. Rev. D}, {\bf 48}:3731, (1993)}.
%
\bibitem{Padmanabhan1996}
Sriramkumar, L., and Padmanabhan, T., Response of ﬁnite time particle detectors in noninertial
frames and curved space-time,
\href{http://dx.doi.org/10.1088/0264-9381/13/8/005}
{{\it Class. Quantum Grav.}, {\bf 13}:2061, (1996)}.
%
\bibitem{Satz2008}
Louko, J. and Satz, A., Transition rate of the Unruh–DeWitt detector in curved spacetime,
\href{http://dx.doi.org/10.1088/0264-9381/25/5/055012}
{{\it Class. Quantum Grav.}, {\bf 25}:055012, (2008)}.
%
\bibitem{Louko2012}
Hodgkinson, L. and Louko, J., How often does the Unruh-DeWitt detector click beyond four dimensions?,
\href{https://doi.org/10.1063/1.4739453}
{{\it J. Math. Phys.}, {\bf 53}:082301, (2012)}.
%
\bibitem{Louko2014}
Juarez-Aubry, B. A., and Louko, J., Onset and decay of the 1 + 1 Hawking–Unruh effect:
what the derivative-coupling detector saw,
\href{https://iopscience.iop.org/article/10.1088/0264-9381/31/24/245007}
{{\it Class. Quantum Grav.}, {\bf 31}:245007, (2014)}.
%
\bibitem{Louko2016}
Fewster, C. J., Juarez-Aubry, B. A., and Louko, J., Waiting for Unruh,
\href{http://dx.doi.org/10.1088/0264-9381/33/16/165003}
{{\it Class. Quantum Grav.}, {\bf 33}:165003, (2016)}.
%
\bibitem{Moustos2019}
Juarez-Aubry, B. A., and Moustos, D., Asymptotic states for stationary Unruh-DeWitt detectors,
\href{https://journals.aps.org/prd/abstract/10.1103/PhysRevD.100.025018}
{{\it Phys. Rev. D}, {\bf 100}:025018, (2019)}.
%
\bibitem{Jenkinson2025}
Dickinson, R., Forshaw, J., Jenkinson, R., and Millington, P., A new study of the Unruh effect,
\href{https://iopscience.iop.org/article/10.1088/1361-6382/ad9c12}
{{\it Classical and Quantum Gravity}, {\bf 42}:025014, (2025)}.
%
\bibitem{Louko2024}
Bunney, C. R. D. and Louko, J., Circular motion in (anti-)de Sitter spacetime:
thermality versus finite size,
\href{https://arxiv.org/abs/2406.17643v1}{arXiv:2406.17643 (2024)}.
%
\bibitem{SuppMat}
See Supplemental Material for more details, which includes Refs. [44-45].
%
\bibitem{Abramowitz1972}
Abramowitz, M. and Stegun, I. A., {\it Handbook of Mathematical Functions} (Dover, New York, 1972).
%
\bibitem{Gradshteyn2007}
Gradshteyn, I. S. and Ryzhik, I. M., {\it Table of Integrals, Series and Products} Seventh Edition,
(Academic Press, San Diego, 2007).
\end{thebibliography}
\end{document}